\documentclass[runningheads]{svmult}

\usepackage{makeidx}   
\usepackage{graphicx}  
\usepackage{subeqnar}  
\usepackage{multicol}  
\usepackage{physprbb}  
\makeindex             

\newcommand{\ltsima}{$\; \buildrel < \over \sim \;$}
\newcommand{\simlt}{\lower.5ex\hbox{\ltsima}}
\newcommand{\gtsima}{$\; \buildrel > \over \sim \;$}
\newcommand{\simgt}{\lower.5ex\hbox{\gtsima}}

\begin{document}
\title*{Chemical Abundances and Milky Way Formation}

\toctitle{Chemical Abundances and Milky Way Formation}
\titlerunning{Chemical Abundances and Milky Way Formation}
\author{Gerry Gilmore\inst{1}
\and Rosemary F.G. Wyse\inst{2}}
\authorrunning{Gilmore \& Wyse}
\institute{ Institute of Astronomy, Madingley Road, Cambridge, UK
\and Department of Physics and Astronomy, Johns Hopkins University,
 Baltimore, MD 21218, USA}

\maketitle              

\begin{abstract}
Stellar chemical element ratios have well-defined systematic trends as
a function of abundance, with an excellent correlation of these trends
with stellar populations defined kinematically. This is remarkable,
and has significant implications for Galactic evolution. The source
function, the stellar Initial Mass Function, must be nearly
invariant with time, place and metallicity. Each forming star
must see a well-mixed mass-averaged IMF yield, implying low star
formation rates, with most star formation in at most a few regions of
similar evolutionary history. These well-established results are
difficult to reconcile with standard hierachical formation models,
which assemble many stellar units: galaxy evolution seems to have been
dominated by gas assembly, with subsequent star formation. Recent
results, and some new ones, on the Galactic bulge, stellar halo and thick disk that 
justify this conclusion are presented.
\end{abstract}

\section{The Context: abundances and galaxy formation}

The $\Lambda$CDM paradigm for structure formation in the Universe,
described in many hundreds of published papers, is very effective at
reproducing observed large scale structure, based on a boundary
condition of a scale-free Gaussian random power spectrum. Yet
$\Lambda$CDM contains no information on the physics of whatever makes
up CDM, and remains deficient in its description of galaxies and
small-scale structures: thus it is on galaxy scales and smaller where
we can still learn the most, and hopefully attach some (astro-)physics
to an {\sl ab initio} power spectrum.

Among the very many studies which emphasise the current progress in
studies of galaxy-scale predictions, we note as recent examples
Glazebrook etal (2004) and Cimatti etal (2004), which highlight the
buildup of massive spheroids at earlier times than predicted; de Blok
(2004) showing CDM mass profiles on small scales are shallower than
predicted; D'Onghia \& Lake (2004), on the substructure problem,
showing it extends from single galaxies to galaxies in groups; and
Abadi etal (2003) illustrating how detailed comparisons of simulations
and data for the Solar neighbourhood are becoming feasible.

How will we identify the extra astrophysics required to reconcile the
properties of CDM dark haloes with those of luminous galaxies? We can
start by developing knowledge of the evolutionary history of at least
one place in at least one galaxy. We would be unlucky if that place
were far from the norm: alternatively, any theory that predicts
such a history to be very unusual might be suspect - the galaxian
Copernican principle. Kinematics and current spatial location are of
course critical parameters, so that traditional stellar populations
analyses are prerequisities. However, kinematical histories, as
represented by velocity dispersions, are at best confused by
phase-wrapping over time, and can be largely lost if a
virialisation process is associated with significant mergers. Kinematics
do provide valuable statistical information, but, apart from the
single (approximately, sometimes) conserved quantity of angular
momentum, provide direct information only about very recent mergers, a
minor aspect of Galaxy formation in any model. The additional
complementary information is best available from chemical abundances,
especially including element ratios.

It is remarkable that chemical abundances are a valid, and relatively
robust, tracer of galactic evolution: it is worth considering why this
is possible.

Figure 1 (from Venn etal 2004) is a compilation of most of the recent
high-quality element ratio data for Galactic field stars (together 
with data for some stars in several satellite galaxies). The remarkable
implication of this figure is that the overwhelming majority of
Galactic field stars have element ratios relative to Fe which have a
scatter which is always relatively small, over a range of some 4 dex
in [Fe/H]. This small scatter is quite contrary to much simple
expectation, and has profound implications. 

\begin{figure}[t!hb]
\begin{center}
\includegraphics[width=.6\textwidth]{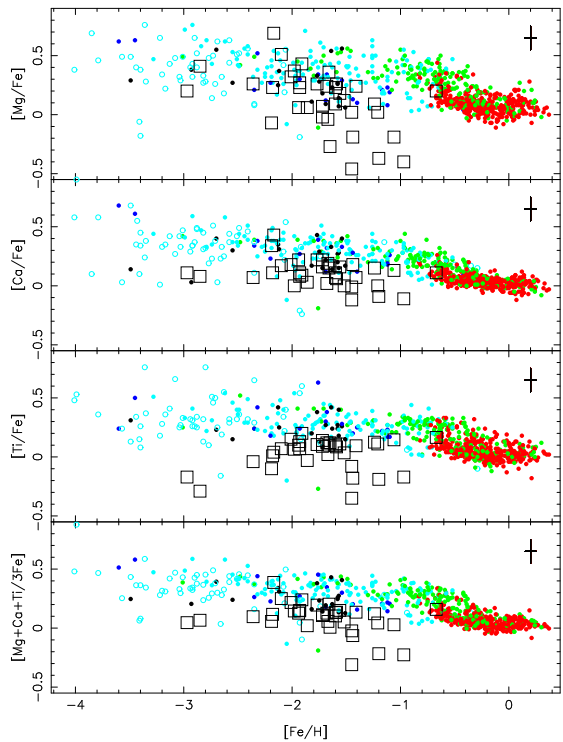}
\end{center}
\caption[]{Element ratio data for Galactic field stars (dots) compiled
by Venn etal (2004). [The open squares are stars in dSph galaxies.] The
important point to notice here is the very small scatter in element abundances
for Galactic field stars at any given [Fe/H] value, over 4dex in [Fe/H].}
\label{gg-1-f1}
\end{figure}

One implication follows from the continuity in the pattern of the
elemental ratios as a function of [Fe/H]: at every metallicity, all
forming stars have element ratios which require that they are being
formed from gas which has a common well-mixed history. At every
metallicity, the star-forming gas must be well-mixed, and only mildly
different in chemical element-ratio enrichment than that gas which
formed the previous stellar generation. This seriously restricts the
role of significant inflows of gas over an extended period of time, as
inflows of gas with a very different enrichment history would induce
scatter. However, inflow of metal-free gas (as often invoked to
`solve' the local disk `G-dwarf problem') would reduce [Fe/H] while
leaving [$\alpha$/Fe] unchanged.  A second implication follows from
the dependence of element production on the initial mass of the
main-sequence progenitor of the supernova.  The observed small
dispersion in element ratios, even under the extreme assumption of
perfect ISM mixing, requires that each new-forming star sees an
approximately invariant, and mass-averaged, IMF (eg Wyse \& Gilmore
1992; Nissen et al.~1994). The rate of star formation must be low
enough to allow time for element creation followed by large-scale
mixing.

That is, the straightforward interpretation of abundance data for
Galactic field stars in terms of stellar populations is feasible only
because the Galaxy apparently acquired its gas early, or at a rate
which was well-matched to the star formation rate across the whole
volume now sampled by local halo stars, and kept this gas well-mixed;
and because the stellar IMF is (close to) invariant over time and
metallicity. Neither deduction was obvious, nor is the underlying
physics understood.  However, these two deductions apply so well they
have become assumed: authors use any violation to rule out some
possible Galaxy merger histories, as in the Venn etal analysis from
which Figure~1 is taken.

We now briefly consider in turn the abundance constraints on the major
Galactic stellar populations, highlighting the poorly known aspects.

\section{The Galactic Bulge}

It is quite astonishing how little observational information is
available on the chemistry and kinematics of the Galactic Bulge. A
massive and exciting opportunity for the current multi-object
spectrographs is being overlooked here. 

The central regions of the Galaxy, including the Bulge, should
accumulate any tightly-bound high phase-space density material
associated with mergers (since the most-bound regions of substructure
end up in the most-bound regions; Zurek, Quinn \& Salmon 1988);
low-angular momentum material in general; material driven in by bar
asymmetries in the gravitational potential; and any accreted gas which
can cool efficiently and lose angular momentum. The earliest stars
formed will be there, even in the merger scenario for build-up of
bulges (e.g. Kauffman 1996), while the whole will be affected in some
as yet undefined way by feedback from the central Super-Massive Black
Hole. It should be a very complex place, well worth detailed
investigation. The luminosity scale height of the Bulge is of order
300pc, so that even the best studied `inner' field, Baade's Window at
coordinates ($\ell = 1$, $b = -4$) or a projected Galactocentric
distance of $\simgt 500$~pc, is itself rather far from the heart of
the matter.

In so far as these properties are determined, the outer bulge, beyond
2 scale lengths, is apparently predominately mildly metal-rich
(perhaps 0.5 Solar in the mean) with a very broad abundance
distribution function, is older than $\sim 10$Gyr, ie is
indistinguishable from the old metal-rich globular clusters, which may
be related, and is alpha-element enhanced (e.g.~Ibata \& Gilmore
1995a, 1995b; Sadler, Rich \& Terndrup 1996; Zoccali et al.~2003;
Fulbright, Rich and McWilliam 2004): this does not favour merger
origin models, but does suggest predominant in-situ star formation. A
star formation rate of $10 M_{\odot} yr^{-1}$ is implied. The angular
momentum distribution function is dominated by very low angular
momentum, and is strongly dissimilar to the corresponding distribution
for the Galactic disk: this argues against a formation process through
disk-bar evaporation (Figure 2, and cf. Ibata \& Gilmore 1995b), and
suggests a close connection between bulge and halo in formation.

\begin{figure}[th]
\begin{center}
\includegraphics[width=.6\textwidth]{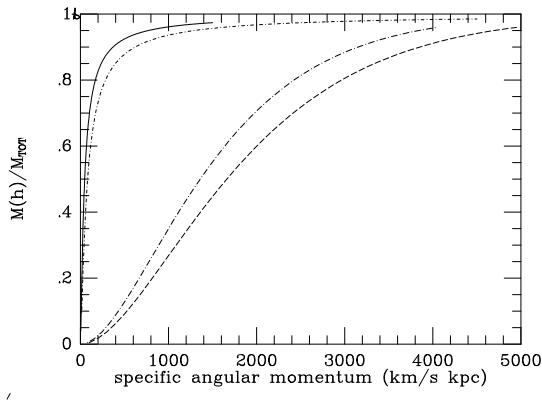}
\end{center}
\caption[]{The distribution of specific angular momentum in the
dominant Galactic stellar populations. This indicates the similarity
in this fundamental parameter, and presumably in origin, between bulge
and halo (the two curves near the top left of the figure), and the
quite distinct bahaviour of the Galactic thin and thick disks (the
two curves through the centre of the figure). This figure is from Wyse
\& Gilmore 1992.}
\label{gg-1-f2}
\end{figure}

The structure of the inner Bulge, where most of the mass is, is
inevitably confused by the inner disk, and by the extremely dense
stellar super-cluster with small scale length concentrated at the very
centre with its associated black hole (Schodel, Ott, Genzel etal
2003). Recent studies in the IR have hinted at the considerable
complexity of the central stellar populations (van Loon, Gilmore,
Omont etal 2003), with young high-mass star and cluster formation at
solar metallicity (Najarro etal 2004), a vast reservoir of molecular
gas to feed continuing star formation (Pierce-Price etal 2000), and
complex bar and disk structure (Bissantz, Debattista \& Gerhard
2004). Extant analyses have usually assumed the bar is associated with
the bulge, rather than with the inner old disk, or, more commonly, do
not make any distinction between these various central populations,
even though they may well have very different kinematic, age and
abundance distribution functions, and have formed in very different
ways.  Unravelling their inter-relationships is not yet possible due
to the limited data.  At present one needs to be careful to define
terms, such as `main bulge' (cf.~Wyse, Gilmore \& Franx 1997)

Considerable progress is currently being made in mapping the very
inner Galaxy. In a recent study, Babusiaux \& Gilmore (2004) have used
near-IR photometry to map the inner galaxy in the Galactic Plane
(figure 3), using red clump stars as tracers. Their photometric work
suggests that the bulge stars 1-deg from the Plane (ie, 4 times closer
in than Baade's Window) are metal-rich and alpha-enhanced, while the
stars which define the bar itself, measured within 0.2deg of the
Plane, best match intermediate-age isochrones without alpha-enhancement, 
The stars in the Bar itself seem more disk-like than those stars
seen in Baade's window and other fields at similar high latitudes. Direct
spectroscopic confirmation, to ensure age-metallicity degeneracies are
not confusing the photometric analyses, should be possible, now that
the individual bar clump giants have been identified.

\begin{figure}[hb]
\begin{center}
\includegraphics[width=.6\textwidth]{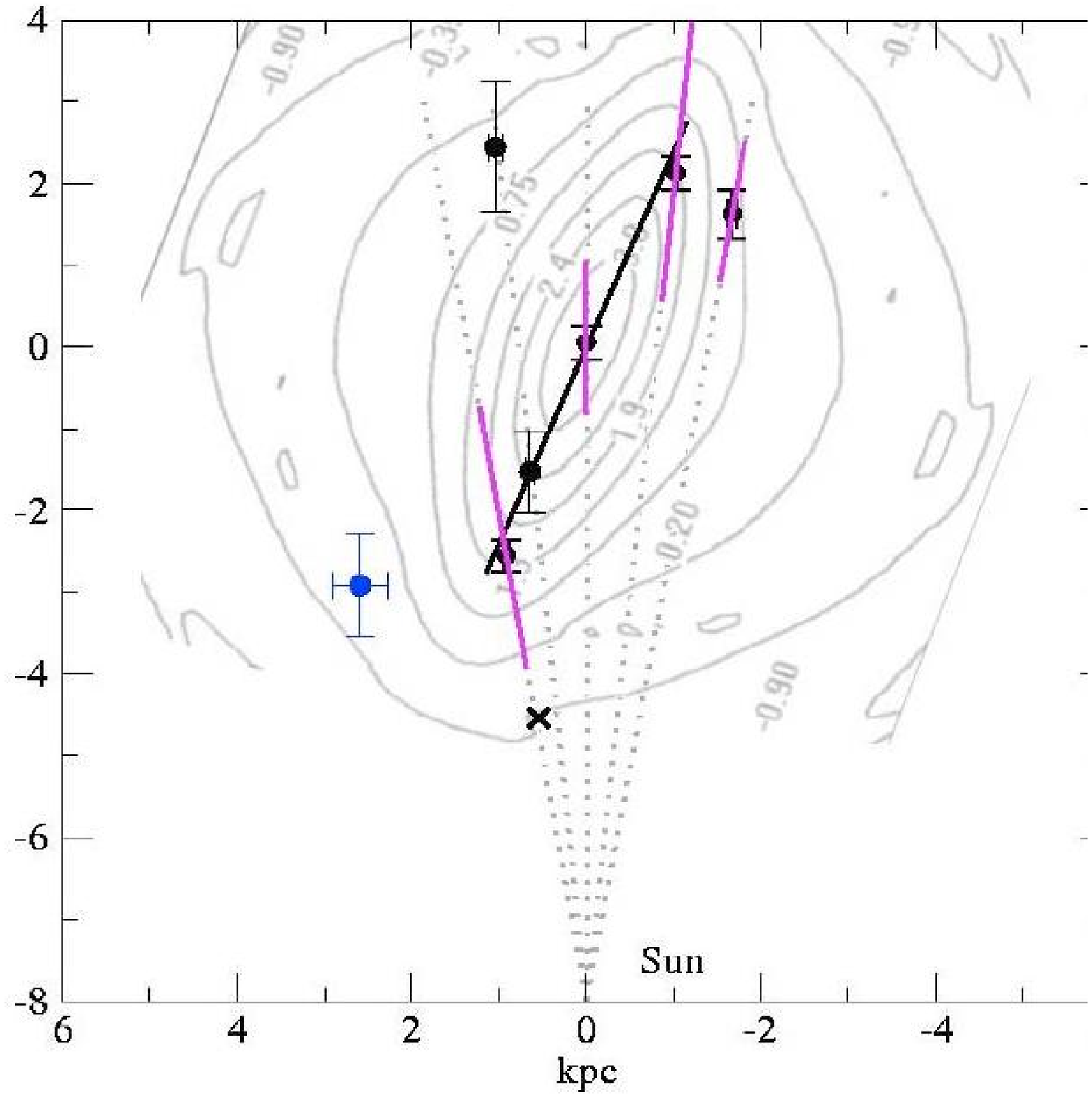}
\end{center}
\caption[]{The 2-D structure of the inner Galactic bar, based on
direct photometric analysis of red clump stars as distance indicators.
Now these individual bar stars are identified, detailed study of their
kinematics and chemical abundances will be possible. Preliminary
photometric indications suggest the bar is more closely disk than
bulge in its populations. This figure is from Babusiaux \& Gilmore 2004.}
\label{gg-1-f3}
\end{figure}

\section{The Stellar Halo}

Figure 2 above illustrates a similarity in at least one fundamental
parameter between the Bulge and the stellar halo. The stellar halo
traced by stars passing near the Sun is some 30\% of its total mass of
$\sim 2 \times 10^9 M_{\odot}$. These stars are predominantly old and
metal poor. They show remarkable uniformity in their relationship
between the element ratios and the total metal abundance (Figure 1),
with implications noted above.

It has been known for many years (eg Unavane et al 1996; Gilmore \&
Wyse 1998), and has been re-emphasised recently (this meeting) based
on much better data, that the stars in the stellar halo are unlike
those in surviving Galactic satellites.  Considerable progress is
being made in determining the properties of the current Galactic
satellites, especially the fundamental properties, mass density and
dark matter profile, as well as in analysing the tracers of star
formation histories, CMDs and chemical abundances.  All the dSph
satellites will have good kinematic maps complete within the near
future, almost all based on FLAMES data, and all providing useful
chemical abundance maps (UMi: Kleyna etal 2003; Draco: Kleyna etal
2001; Sextans: Kleyna etal 2004; LeoI, LeoII, Carina: Harbeck etal this
meeting; Gilmore etal in prepn; Sgr: Ibata, Wyse, Gilmore etal 1997;
Scl: Tolstoy etal, this meeting; Fornax: Tolstoy etal in prepn). All
have similarities and complexities, yet none provides a source for a
`typical' field halo star. The different patterns of elemental
abundances in the dwarf galaxies, which apparently rules them out as
`building blocks' for the Milky Way, can be understood in terms of
their different star formation histories (cf.~Gilmore \& Wyse 1991).

This suggests strongly that the stellar halo formed in place early in
Galactic history - a conclusion apparently required by the data, yet
strongly in disagreement with most current galaxy formation
simulations. The simulations imply significant late accretion of
galaxies into the Milky Way: do we see this? A discovery which
encouraged much current enthusiasm was that of the Sgr dwarf (Ibata,
Gilmore \& Irwin 1994, 1995): it is salutary to recall that Sgr was
recognised as peculiar in real time at the AAT (during the
Ibata/Gilmore bulge study) because both its velocity and its stellar
populations were quite unlike any expected (or observed) field
star. Sgr does however single-handedly make a substantial perturbation
to the stellar mass (and age mix) of the stellar halo. More recently a
second, and possibly a third, structure has been identified in the
outer halo from SDSS photometry (Newberg et al.~2002); their
relationship to the Sgr dwarf is unclear, but most confirmed structure
may be ascribed to this galaxy (cf.~Majewski et al.~2003).  A firm
conclusion is that in another Hubble time, the stellar halo will be
relatively younger than it is today (Wyse 1996).

The halo globular cluster system also provides valuable information,
since accurate distances, and hence reliable ages, can be
derived. Mackey \& Gilmore (2004) recently acquired and compiled a new,
nearly complete, internally consistent set of photometric studies
of the globular cluster population in both the Milky Way and its
satellite galaxies: they deduce, from analysis of HB morphology, age,
abundance and structural information, somewhat more relaxed limits on
accretion of the halo than earlier analyses, but still preclude late
accretion as a dominant factor. 

Study of the relative importances of assembly of stellar systems and
in situ star formation  remains extremely active,
and will presumably soon converge as the large area CCD photometric
surveys cover the whole sky, and accurate spectroscopic  surveys
begin to acquire large samples.

\subsection{M31 and its halo/bulge/thick disk}

An interesting case here is M31, which has long been suspected of
having tidally stripped M32 (Faber 1973; Choi, Guhathakurta \& Johnson
2002), and now we have kinematic evidence that it is also currently
accreting from NGC~205 (McConnachie et al.~2004). However, the
equivalent of the Milky Way's halo population II, a metal-poor
component to the `halo', traced by RR Lyrae stars and by metal-poor
giants, is not dominant even at very many photometric scale lengths
from the centre of M31 (e.g.~Durrell, Harris \& Pritchet 2004). The
relation to accreted sateliites is clearly complex (e.g.~Ferguson et
al.~2002). The dominant non-thin-disk population in M31 is rather
metal-rich, with a mean of $\sim -0.6$~dex (Mould \& Kristian 1986;
Durrell et al.~and references therein). Wyse \& Gilmore (1988) discuss
the apparent equivalence between this extended intermediate population
in M31 and the Galactic thick disk, an equivalence which has become
stronger as studies of M31 have developed (Sarajedini \& van Duyne
2001; Brown et al.~2003). A fuller interpretation in terms of the mass
assembly history of M31 awaits the current wide-area coverage with a
combination of deep CMDs for age constraints, together with
spectroscopic kinematics and metallicities.

\section{The Old Disks}

The thick disk probably was caused by, and/or is the remnant of, one
or more (early?) merger events. The (local) mean thick disk metallicity
of -0.6 dex and its old age suggest a very massive satellite
-- comparable at least to the LMC? -- was destroyed if the (local) thick disk is
its remnant.  Simulations (eg Abadi etal 2003) suggest at
least the old thick disk will be the remnant of several merged
sub-galaxies, with the relative mix changing with radius. This may be
consistent with recent AAT 2dF survey results (Gilmore, Wyse \& Norris
2002) which suggest that the distribution function of angular
momentum, and perhaps metallicity, a few kpc from the Plane is not
consistent with simple local extrapolations.

One of the more challenging things to understand in these
multiple-fragment models is the evolution of the present-day surviving
thin old disk, particularly since a recent accretion of the thick disk
could destroy the thin disk.  Further, in these models stars should
not form in a thin disk at all until after most of the merging is
complete, since otherwise disks have too small a scale-length (Navarro
\& Steinmetz 1997). A solution to both these issues may be that the
local old thin disk is also accreted in these models.  Here the recent
results of many authors (cf Figure 1 above) have shown there is a very
small scatter in element ratios at any [Fe/H] value, particularly
within a stellar population, yet there is an extremely large scatter
in the age-[Fe/H] relation at every age (Nordstrom et
al.~2004). Linking the local and global remains a challenge.


\begin{thebibliography}{8.}
\addcontentsline{toc}{section}{References}

\bibitem{} Abadi, M., Navarro, J., Steinmetz, M. \& Eke, V. 2003  ApJ
597 21 


\bibitem{} Babusiaux, C., \& Gilmore, G. 2004 MNRAS submitted

\bibitem{} Bissantz, N., Debattista, V. \& Gerhard, O. 
2004 ApJ 601 L155

\bibitem{} Brown, T. et al.~2003 ApJL 592 L17
\bibitem{} Choi, P., Guhathakurta, P. \& Johnson, K. 2002 AJ 124, 310

\bibitem{} Cimatti. A., Daddi, E., Renzini, A., Vanzella, E., et al.~2004  Nature 430 184

\bibitem{} de Blok, W.J.K. 2004, IAU Symp 220 eds Ryder et al.~(ASP, San Francisco) p69

\bibitem{} D'Onghia, E., \& Lake, G., 2004 ApJ 612 628
\bibitem{} Durrell, P., Harris, W. \& Pritchet, C. 2004 AJ 128 260
\bibitem{} Faber, S.M. 1973 ApJ 179 423
\bibitem{} Ferguson, A. et al.~ 2002 AJ 124 1452
\bibitem{} Fulbright, J., Rich, R.M. \& McWilliam, A., 2004 astroph-0411041

\bibitem{} Gilmore, G., Wyse, R.F.G. 1998 AJ  116 748 

\bibitem{} Gilmore, G., Wyse, R.F.G. \& Norris, J.E. 2002 ApJL 574 L39

\bibitem{} Glazebrook, K., Abraham, R., McCarthy, P. et al.~2004 Nature 430 181

\bibitem{} Ibata, R., Gilmore, G. \& Irwin, M. 1994 Nature 370 194


\bibitem{} Ibata, R. \& Gilmore, G. 1995a MNRAS 275, 591

\bibitem{} Ibata, R. \& Gilmore, G. 1995b MNRAS 275, 605

\bibitem{} Ibata, R., Gilmore, G. \& Irwin, M. 1995 MNRAS 277 781

\bibitem{} Ibata, R., Wyse, R.F.G., Gilmore, G., Suntzeff, N. \& Irwin, M. 1997 AJ 113 634

\bibitem{} Ibata, R. et al.~2004 MNRAS 351  117
\bibitem{} Kleyna, J., Wilkinson, M., Evans, N.W. \& Gilmore, G.  2001 ApJ 563
L115

\bibitem{} Kleyna, J., Wilkinson, M., Evans, N.W. \& Gilmore, G. 2004
astroph-0409066

\bibitem{} Kleyna, J., Wilkinson, M., Gilmore, G. \& Evans, N.W. 2003 ApJ 588
L21
\bibitem{} Kauffmann, G. 1996 MNRAS  281 487

\bibitem{} van Loon, J., Gilmore, G., Omont, A.  et al.~2003 MNRAS 338 857
\bibitem{} Mackey, A.D. \& Gilmore, G. 2004 MNRAS 354  470 
\bibitem{} Majewski, S., Strutskie, M., Weinberg, M. \& Ostheimer, J. 2003 ApJ 599 1082
\bibitem{} McConnachie, A. et al.~2004 MNRAS 351  L94 
\bibitem{} Mould, J. \& Kristian, J. 1986 ApJ 305 591

\bibitem{} Najarro, F., Figer, D., Hillier, D. \& Kudritzki, R.  2004 ApJ 611 L105
\bibitem{} Navarro, J. \& Steinmetz, M. 1997 ApJ 478  13

\bibitem{} Nissen, P. E., Gustafsson, B., Edvardsson, B., Gilmore,
G. 1994 A\&A  285 440 
\bibitem{} Newberg, H., Yanny, B., Rockosi, C., etal  2002 ApJ 569 245
\bibitem{} Nordstrom, B., Mayor, M., Andersen, J. etal.~2004 A\&A 418 989 
\bibitem{} Pierce-Price, D., et al.~2000 ApJ 545 L121
\bibitem{} Reitzel, D., Guhathakurta, P. \& Rich, R.M. 2004 AJ 127 2133

\bibitem{} Sadler, E., Rich, R.M. \& Terndrup, D. 1996, AJ 112 171
\bibitem{} Sarajedini, A. \& van Duyne, J. 2001, AJ 122 2444

\bibitem{} Schodel, R., Ott, T.,  Genzel, R. et al.~2003 ApJ 596 1015

\bibitem{} Unavane, M., Wyse, R.F.G. \& Gilmore, G. 1996 MNRAS 278 727
\bibitem{} Wyse, R.F.G. 1996 ASP Conf series 88, eds Trimble \& Reisenegger p128
\bibitem{} Wyse, R.F.G. \& Gilmore, G. 1988 AJ 95 1404
\bibitem{} Wyse, R.F.G. \& Gilmore, G. 1992 AJ 104 144
\bibitem{} Wyse, R.F.G., Gilmore, G. \& Franx, M. 1997 ARAA 35 637
\bibitem{} Zoccali, M.~et al. 2003 A\&A 399 931

\bibitem{} Zurek, W., Quinn, P. \& Salmon, J. 1988 ApJ 330 519

\end{thebibliography}
\end{document}